\documentclass[12pt]{article}
\usepackage[final]{pdfpages}
\usepackage{graphicx} 
\usepackage{amsmath}
\usepackage{amssymb}
\usepackage{epsfig}
\usepackage{cite}
\usepackage{color,colordvi}
\newcommand{\eq}{\begin{equation}}
\newcommand{\eqx}{\end{equation}}
\newcommand{\eqn}{\begin{eqnarray}}
\newcommand{\bi}{\begin{itemize}}
\newcommand{\eqnx}{\end{eqnarray}}
\newcommand{\ei}{\end{itemize}}

\newcounter{hran}

\def\MSbar{\relax\ifmmode\overline{\rm MS}\else{$\overline{\rm MS}${ }}\fi}

\begin{document}

\begin{center}


{\Large\bf  Black Hole's Quantum $N$-Portrait}

\vspace{.1cm}

\end{center}

\begin{center}

{\bf Gia Dvali}$^{a,b,d,c}$ and {\bf Cesar Gomez}$^{e}$


\vspace{.1truecm}

{\em $^a$Arnold Sommerfeld Center for Theoretical Physics\\
Department f\"ur Physik, Ludwig-Maximilians-Universit\"at M\"unchen\\
Theresienstr.~37, 80333 M\"unchen, Germany}


{\em $^b$Max-Planck-Institut f\"ur Physik\\
F\"ohringer Ring 6, 80805 M\"unchen, Germany}


{\em $^c$CERN,
Theory Division,  1211 Geneva 23, Switzerland}


{\em $^d$CCPP, Department of Physics, New York University\\
4 Washington Place, New York, NY 10003, USA}


{\em $^e$
Instituto de F\'{\i}sica Te\'orica UAM-CSIC, C-XVI \\
Universidad Aut\'onoma de Madrid,
Cantoblanco, 28049 Madrid, Spain}\\

\end{center}


\centerline{\bf Abstract}

\noindent

\small{
 We establish a {\it quantum}  measure of  classicality in the form of   the occupation number, $N$, of gravitons in a  gravitational field.   This  allows us to view classical background geometries as 
 quantum Bose-condensates with large occupation numbers of soft gravitons.  
  We show that  among all possible sources of a given physical length,  $N$ is maximized by the black hole  and coincides with its entropy.  
  The emerging quantum mechanical picture of a black hole is surprisingly  simple and fully 
parameterized  by  $N$.   
   The  black hole  
 is a leaky bound-state in form of a cold Bose-condensate of  $N$ weakly-interacting soft gravitons of wave-length 
 $ \sqrt{N}$ times the Planck length   and of quantum interaction strength $1/N$.   
 Such a bound-state  exists  for  an arbitrary $N$. 
   This picture 
     provides a  simple quantum description of the phenomena of Hawking radiation,  Bekenstein entropy as well as of non-Wilsonian UV-self-completion of Einstein gravity. 
   We show  that  Hawking radiation is nothing but a quantum depletion of the graviton Bose-condensate, which despite the zero temperature  of the condensate produces a thermal spectrum
 of temperature $T \, = \, 1/\sqrt{N}$.   
  The Bekenstein entropy originates from the exponentially growing with $N$ number of  quantum 
states.   Finally, our quantum picture  allows to understand classicalization of deep-UV gravitational scattering   as  $2 \rightarrow N$ transition. 
     We  point out some fundamental  similarities between the black holes and 
  solitons,  such as a t'Hooft-Polyakov monopole.  Both 
   objects  represent Bose-condensates of  $N$ soft bosons of wavelength  $\sqrt{N}$
   and interaction strength $1/N$.   
   In short, the semi-classical black hole physics is  $1/N$-coupled  large-$N$ quantum physics. }

\vskip .4in
\noindent

\newpage


\section{Introduction and Summary} 

Einstein's  general theory of relativity (GR) is a classical theory of gravity. Viewed as a quantum theory it is a theory that 
propagates a unique  weakly-coupled quantum particle  with zero mass and  spin-2. 
 At low energies,  a dimensionless quantum self-coupling of gravitons can be consistently 
 defined as,  
 \begin{equation}
 \alpha_{gr} \, \equiv \, \hbar  G_N \lambda^{-2}  \, .
 \label{alphag}
 \end{equation}
Or equivalently, $\alpha_{gr}$ can be rewritten as the ratio of the two length-scales, 
the {\it quantum}  Planck length  ($L_P \, \equiv \, \sqrt{ \hbar G_N}$ )  
and the {\it classical}  wave-length  $\lambda$, 
\begin{equation}
\alpha_{gr} \, \equiv  \,  {L_P^2  \over \lambda^2} \, .
\label{alphagp}
\end{equation}
For wavelengths $\lambda \, \ll\, L_P$, the above coupling becomes strong, and the theory violates perturbative  ( in $\alpha_{gr}$) unitarity in graviton-graviton scattering.  The standard approach to the issue is to assume that 
Einstein gravity requires a UV-completion by integrating-in some weakly-coupled new physics 
at some  distance  $  >  \, L_P$ exceeding the perturbative-unitarity breakdown scale.  
 The hope then is that such new physics would restore perturbative unitarity in sub-Planck-length 
 region and shall enable us to  compute  gravitational amplitudes at arbitrarily short distances.  

  In \cite{gia-cesar} we have suggested an alternative root, which abandons the  possibility of any kind of 
  Wilsonian UV-completion. Instead the idea is that Einstein gravity viewed as a quantum field theory  is {\it self-complete}  in a non-Wilsonian way.   The theory  prevents us from probing  shorter than $L_P$ distances by responding to any high-energy scattering  by  producing 
 large occupation number $N$ of very long wave-length 
  $\lambda  \,  \gg \, L_P$ particles.   Unitarity is restored,  because  the processes in which momentum  transfer per particle exceeds  $\hbar / L_P$ are extremely rare  and appear with exponentially-suppressed probability.  
   
     In \cite{class} the above  concept of non-Wilsonian UV-completion was generalized to other classes of theories  and was termed  {\it classicalization}. 
    
     One of the consistency tests for  non-Wilsonian quantum-self-completeness of Einstein gravity is  to give a quantum description of the  black hole physics. It is sometimes believed that the explanation of seemingly-mysterious black hole properties, such as thermality and Bekenstein entropy, is impossible within the existing framework of Einsteinian gravity, even if we treat it as a quantum theory. 
   The  goal of this paper is to provide the opposite point  of view. 
  We shall show that Einstein's gravity when viewed as a quantum theory of gravitons can 
  account for all the above  black hole properties  in terms of familiar phenomena  very well understood in the 
  ordinary quantum mechanical setting.   
  
  In order to do this, we need to understand what is the meaning of classicality from the quantum point of view. 
  Black holes usually are described in the language  of classical solutions and geometry. 
   But  nature is quantum.   In our framework, any classical object is understood as a quantum bound-state of high occupation number, $N \, \gg \, 1$.   Such bound-states generically may have many characteristics. Such as,  for example, different relevant wave-lengths  and different oscillation patterns and frequencies.    As we shall show, among the variety of macroscopic objects, black holes and solitons stand out  as the simplest ones.  Their physics can  be described by a   single quantum characteristic, $N$.   We shall see, that all the above properties of black holes can be understood in terms of this single number $N$. 
 This number for us serves as  the main measure of classicality.

  The concept of  $N$, (the occupation number of bosons produced in an arbitrary classical field)
     as the measure of classicality was introduced in   \cite{gia-gomez-alex}. 
 As shown there,  in terms of familiar classical characteristics of  gravitating sources this number is equal to a total mass of the source $M$ times its 
 gravitational radius $r_g$ divided by $\hbar$, 
 \begin{equation}
   N \, = \, Mr_g/\hbar  \, ,
 \label{master}
 \end{equation}
and is insensitive to other characteristics. 
  The present paper is about  explaining the power and the physical significance of this number
  and its role in shedding  a new light on the fundamental properties of black holes, such as 
  thermality and entropy.   Our goal is to show, that quantum black hole physics can be 
  formulated in terms of $N$. In particular the familiar geometric and semi-classical characteristics 
  of black holes, such as horizon and temperature, are fully-emergent  approximate 
  notions that become the useful language  in  large $N$ limit.

  Most importantly, $N$ is a {\it universal} occupation number of gravitons irrespective
  of the particular nature of the source.  
   Among all possible sources of some characteristic  physical size $R$, the number $N$ is maximized by a black hole. That is,  for maximal $N$ the wavelength of the occupying quanta 
$\lambda$  saturates the length that classically would be defined as the gravitational 
Schwarzschild radius  of the configuration.    Or equivalently  $N$ is maximized  for     
  $\lambda \, = \, R \, = \, r_g$.  From this perspective, black holes are the most classical object among all possible objects of a given characteristic wave-length $\lambda$. 
   
    Universality of $N$ leads us to a quantum-mechanical picture of a black hole 
  which does not rely on the classical geometric notions. Instead it is 
  fully characterized by the single parameter $N$. 
    
    In our description a black hole is  a {\it leaky} bound-state, a Bose-condensate of $N$ weakly-interacting soft gravitons  of the wave-length determined by $N$ as,
  \begin{equation}
   \lambda \, = \, \sqrt{N}  L_P  \, .
    \label{lambdaN}
    \end{equation}   
    The interactions among individual constituent  gravitons  are  weak,  and given by  an effective dimensionless gravitational coupling  $\alpha_{gr}$.  The wavelength dependence 
     of this coupling given by (\ref{alphag}),  fixes,   
    \begin{equation}
     \alpha_{gr} \, = \, 1/N \, . 
     \label{weak-coupling}
     \end{equation}
   Each individual graviton is subject to a collective binding potential created by  $N$-quanta 
   of the range $\sqrt{N} L_P$ and the  strength 
 \begin{equation}
 V \, = \,  {\hbar  \over   \sqrt{N} L_P}  
  \label{potential1}
  \end{equation}
  
  As we see,  in our quantum picture $N$ is the sole characteristic of the black hole.  It is 
 an intrinsically-quantum parameter and  replaces the classical notions of  the black hole mass and the gravitational radius. 
 Since $N$ can take arbitrarily large values,  this  immediately  explains 
 why classically the  black hole mass appears as an arbitrary integration constant. 
 It also explains why a naive quantum intuition of thinking about black holes as a box 
 filled with particles of fixed energy (or temperature) is completely wrong. 
  
   In reality black holes are self-imposed bound-states in which the number of particles, their wave-lengths and binding energy are strictly correlated  via equations (\ref{lambdaN}) 
 and (\ref{potential1}). 
   
    Thus, a black hole is a Bose-condensate of  $N$ weakly-interacting 
    (via coupling $1/N$) gravitons of wavelength  $\sqrt{N}L_P$ in which for any $N$ the escape  energy  is just above the energy of the condensed gravitons. 
   Hence the black hole is a {\it leaky-condensate}.   We shall see that,  despite the fact that  the condensate has a zero temperature,  the thermal spectrum of Hawking radiation  of temperature
   \begin{equation}
     T \, = \,  { \hbar \over \sqrt{N} L_P}
     \label{temperature1}
     \end{equation} 
   emerges as a quantum  effect due to the quantum leakage  of the condensate. 
 

  We shall show that the above picture leads us to an incredibly simple explanation of 
  the following phenomena: 

\begin{itemize}  

 \item  a)  Hawking radiation;

 \item  b)  Bekenstein  entropy of a black hole;

  \item c)  Self-UV-completion of gravity by classicalization \cite{gia-cesar}
  
  \end{itemize}

  Let us briefly summarize  each of these three items.   
  
  $~~~~$

 {\bf a) The Emergent Thermality: }   In our picture, Hawking radiation naturally emerges as  a quantum process in which a single graviton gains an above-threshold 
energy ($ > \, \hbar / \lambda$)  by scattering off the collective potential.  
   This gives the evaporation rate of a black hole described by the equation
 \begin{equation}
    {d N \over dt} \, = \, - \, {1 \over \sqrt{N} L_P } \, , 
\label{rate}
\end{equation}
and the half-life time 
 \begin{equation}
     \tau \, = \, N^{3/2} L_P  \, .
\label{time}
\end{equation}   
   This quantum picture reproduces Hawking's semiclassical  result without any reference to geometric concepts  such as horizon.
  The temperature (\ref{temperature1}) is an emergent notion as a result of quantum depletion 
  of a cold Bose-condensate.   The negative heat-capacity is a simple consequence of the 
  decrease of  $N$ during this depletion.

  
  $~~~$

 {\bf b) Emergence of   Bekenstein  Entropy:}  We show that   Bekenstein entropy emerges as a natural degeneracy of   $N$-graviton states  that scales exponentially with $N$.      
   
   $~~~$

 {\bf c)  Self-Completion by Classicalization: }  The self-completion of gravity by classicalization can be understood as a necessary evolution of any initial 
  two-particle (or few-particle) state of trans-Planckian center of mass energy  into a state with $N$-gravitons of degeneracy  $e^N$.

  $~~~$

 In the present paper  we shall also  point out a fundamental  similarity and complementarity between properties  of black holes and solitons in weakly-coupled gauge theories and/or 
 baryons \cite{Witten} in QCD with large number of colors. 
 
  The solitons are known object  that  {\it classically}  represent solutions of the equations of motion  with a well-defined size  $R$.  
 However, quantum mechanically   the meaning of   $R$  is of the wave-length that corresponds to 
 a maximal occupation number $N$ of the weakly-coupled bosons in the condensate.  
  We shall show that  the quantum physics of solitons can be  formulated in the language of 
$N$ with the emerging picture remarkably similar to the black hole case. 
 
  
   Consequently,  we discover that the quantum physics of black holes and solitons can be understood in the universal 
  language of a single quantum quantity  $N$.  
  Our quantum portraits of these two objects are surprisingly similar. 
  Both,  black holes  as well as solitons represent bound-states  (Bose-condensates)  of  soft quanta 
  with the following characteristics (in units of the  fundamental length):

   \begin{itemize} 
   
   \item    Occupation number  $=\,  N$ 
   
   \item    Wave-length $= \,  \sqrt{N}$
  
  \item  Coupling stength $=  \,  1/ N$
  
  \item  Mass   $= \,  \sqrt{N}$
  
 \end{itemize}

  
   The difference is that, unlike black holes,  solitons (or baryons) exist only for a fixed  $N$,  since the   value is set by the gauge coupling constant as $N  \,  =  \, \alpha_{gauge}^{-1}\,  \equiv  \,  1/(g^2\hbar)$.  Whereas the black hole bound-states 
   exist for arbitrary $N$, because the gravitational coupling  (\ref{alphagp}) is wave-length dependent  and the bound-state condition can be satisfied for arbitrary $N$ as long as 
   the wavelength scales as $\sqrt{N}$.   This peculiarity of gravitational coupling is the main reason why quantum-mechanically   black holes are  {\it leaky} and can deplete, whereas 
   solitons cannot.

   
   

     The  underlying connection through $N$  between the black holes and solitons illustrate why they can be continuously 
    deformed into one another by changing the parameters of the theory.  
    By changing the strength of the gauge coupling, we can move from the solitonic phase, 
   in which the  occupation number is dominated by the gauge bosons, into the black hole phase, in which this number is dominated by gravitons. 
    At the cross-over, to which we refer as the  gauge-gravity transition,  the occupation number 
    of gauge bosons becomes equal to the one of gravitons and we get an extremal black hole.   
   As we shall see, many known classical properties of the charged black holes 
   can be easily understood in this language. 
     
      Our findings give us a  strong evidence that Einstein gravity is self-sufficient  for providing a  fully quantum  picture of the black hole entropy and radiation.  
      Moreover, this picture is remarkably  simple and is described by a single parameter $N$. 
   In this respect, the emerging quantum physics  of black holes appears to be much simpler 
   than the one of other macroscopic bound-states.

    Summarizing our picture in one sentence, we can say that the quantum physics of a  black hole  
    is  large-$N$ physics, which is provided by  the theory for free,  due to energy-self sourcing. 
     In other words,  due to  energy self-sourcing, the number  $N$   is not an input parameter of the theory,  which is  chosen to be 
   large for convenience.  Rather,  it 
   can assume an arbitrarily-large value within the same theory, and  the particle wave-lengths and the interactions strengths are self-tuned  to it  \footnote{Notice, that already for the black holes as light as tens of the Planck  mass,  the smallness of  the $1/N$-expansion parameter becomes 
  comparable  to the electromagnetic  fine structure constant, $\alpha_{EM}$, whereas for an 
  earth mass black hole it becomes of order $10^{-66}$.}.    All  the fundamental differences between the black holes and other large-$N$ systems (such as solitons, or  baryons in QCD with many colors)  originate from this fact.  
   
     The pattern that we are observing over and over in this paper is  that maximally-packed 
  Bose-condensates tend to become maximally simple, and the  black holes  represent the 
  culmination of this tendency.   Extending our approach, of viewing a classical background geometry as a quantum Bose-condensate of the soft gravitons with large occupation number $N$,  to the maximally-symmetric   space-times, we observe that  the analogous $N$-portraits of  de Sitter  and anti de Sitter spaces are  similar to black holes and solitons.  This phenomenon of simplification of the system in an overpacked limit  can be viewed as an underlying quantum  reason for the  holography.   
      
       The  paper is organized as follows. 
       First we slowly  collect a necessary  evidence that leads us to our formulation of 
   black hole's quantum picture in terms of $N$.  Then  we show how  in this picture 
  the familiar  (semi)classical  properties, such as  temperature and entropy emerge.  
  Next we  develop an unified quantum picture  of black holes and solitons  and explain both the fundamental  similarities and differences. 
   Finally, we shall discuss how this picture accounts for self-UV-completion in a non-Wilsonian way  by classicalization.  
   
    Throughout the paper, we shall set the speed of light to one, and we shall dismiss  
all the irrelevant numerical proportionality factors  of order one,  in order to avoid 
unnecessary  defocusing  of the attention.

\section{Classicality in Gravity} 
 The  classical GR contains  no intrinsic length-scale. 
  The only existing parameter in the theory, the Newton's constant $G_N$, has a dimension of length 
  divided by mass, 
  \begin{equation}
             [ G_N]  \, = \, {[length] \over [mass]} \, .
             \label{newton}
  \end{equation}
  So in order to obtain something with dimension of length we need to multiply 
  $G_N$ by something with dimension of mass (or energy).  In the presence of any form of gravitating source,  such a quantity is provided by the energy (mass) of the source, $M$.  Then, combining these two quantities, we can now form a length-scale, 
  \begin{equation}
  r_g \, \equiv \, 2 \,  G_N \,  M \, ,
  \label{gradius}
  \end{equation}
 which represents the gravitational (Schwarzschild) radius of the source of mass $M$.
 In classical gravity $r_g$ is the most important characteristics of the gravitational properties 
 of the source.   The physical meaning of $r_g$ is to set the distance at which the gravitational effect  of a localized source becomes strong. 
 
 It is very important  to note that Schwarzschild radius represents  an intrinsically-classical length, 
 independent of Planck's constant, $\hbar$.  Despite the fact that $r_g$ can be defined for an arbitrary energy source, 
 not every form of energy (e.g., an electron) describes a classical gravitating object.   
 
 In order for the source to be treated as classical, the corresponding Schwarzschild radius 
 must be much larger than the appropriate quantum length-scales in the problem.  
 
  So, what is classicality? 
  
  In order to understand this, we have to note that because of the existence of 
 $\hbar$,  which has dimension of $[mass][length]$, we can form another length-scale, 
 the Planck length, $L_P$, defined from the following equation, 
 \begin{equation}
  L_P^2 \, \equiv  \, \hbar \, G_N \, .
  \label{lp}
  \end{equation}
  Unlike $r_g$, the Planck length is an {\it intrinsically-quantum}  notion. 
 In particular, it  vanishes in the limit $\hbar \, \rightarrow \, 0$.  
 The physical meaning of it can be understood as the length-scale at which the quantum fluctuations of the space-time metric become important and cannot be ignored any longer.  This quantum  meaning 
 will become more precise in due course.  
 
  Analogously, by combining $G_N$ and $\hbar$ we can form a mass scale, the celebrated Planck mass,
  \begin{equation}
  M_P^2 \, \equiv  \, \hbar \, G_N^{-1} \, .
  \label{mp}
  \end{equation}
  
  Other quantum length-scales in the problem  are the  well known Compton  (or de Broglie) 
 wave-lengths of the source of mass (momentum) $M$ ($p$),  
 \begin{equation}
 L_C \, \equiv \,  {\hbar \over M} \,  ~~  {\rm or} ~~ L_{dB}\, \equiv \,  {\hbar \over p}  \, .
 \label{compton}     
 \end{equation}
 The physical meaning of $L_C$ ($L_{dB}$)  is to set the scale at which energy of quantum fluctuations 
 ($E \, = \, \hbar /L_C$)
 would become comparable to the energy of the source.

 Both lengths, $L_P$ and $L_C$ as well as the Planck mass $M_P$, vanish in the limit $\hbar \rightarrow 0 $ when $G_N$ and $M$ are kept fixed.  
 One of the consequences of this fact is that in classical GR 
 ($\hbar\, = \, 0$) one can have black holes of arbitrarily small size.  In reality however 
 $\hbar$ and $G_N$ are the fixed constant and the only parameter we can vary is the mass of the source $M$.   The classicality is then achieved when we  increase $M$ so that 
$r_g \,  \gg \, L_P, L_C$.   
    
  Consider now a gravitating source of a rest mass $M$, for which $r_g \, \gg\, \, L_P, L_C$. 
  Approaching this source from infinity,  we shall encounter strong classical gravitational effects
 way before the quantum effects become important.  Intuitively it is clear that such a source 
 should be described as a classical gravitating object. 
 
  But,  what is the precise quantum field-theoretic meaning of classicality?
  
  Following \cite{gia-gomez-alex},  we shall now introduce the concept of the occupation number of gravitons $N$, as the parameter that measures the level of classicality.    In order to understand this,  let us again consider a gravitating source of 
mass $M$.  The precise quantum composition of this source is unimportant for the present consideration.  This source can equally well represent a collapsing two-particle state  
of center of mass energy $M$ or a soliton of the same mass.  We shall resolve the composition of the source later, but for a moment we shall leave it unspecified. 

 For definiteness, we can  take the source to be spherical of uniform density, and 
of a physical radius $R$ well above  of its gravitational radius $ R \, >> \, r_g \,$.    
  For such a source, the  approximation of linear gravity is valid everywhere, and the gravitational field produced by  it can be easily found. For example, the Newtonian component of the metric perturbation about the flat space outside the source  
    is given by the  well-known $1/r$-potential, 
\begin{equation}
  \phi (r)  \, = \, - \,  {r_g \over r}  \, , 
  \label{field}
  \end{equation}
 and falls-off as $r^2$ for  $r < R$.  
  From the quantum field theory point of view, the above linearized metric represents a superposition 
  of gravitons.  The level of classicality is measured by their occupation number. 
   These gravitons are non-propagating longitudinal gravitons, but this is unimportant 
  for characterizing the classical properties of the fields. 
    We can think of these gravitons as representing  a Bose-condensate.  The only peculiarity 
    of this condensate is that, as long as  $R \, \gg \, r_g$, the condensate cannot self-sustain. 
   In order  to exist,  it requires an external source.  The situation however changes dramatically 
   once $R$ crosses over beyond $r_g$. At this point the condensate becomes self-sustained. 
    This state of a condensate of quantum particles is what we classically call a black hole. 
   This is a key ingredient of our black hole quantum portrait. 
   
   Let us however get there slowly and for the time being stay in the regime  $R \, \gg \, r_g$.  
The measure of classicality of this field is the occupation number of gravitons in it, $N$. 
    This number can be found easily by the Fourier analysis of the 
   metric perturbation (\ref{field}) and is  equal to 
   \begin{equation} 
      N \, = \,  {1 \over \hbar}  M r_g \, .
      \label{N}
      \end{equation}
  The physical meaning of  the number  $N$  becomes transparent from the following reasoning.  
 The gravitational part of the energy is,  
\begin{equation}
E_{grav} \, \sim \  {M r_g \over  R} \, .
\label{Eg}
\end{equation}
We should think of this energy as being the sum of  the energies of the individual gravitons with 
the wave-lengths $\lambda$ and the occupation numbers $N_{\lambda}$, 
\begin{equation}
E_{grav} \,  \sim \,  \sum_{\lambda} \, N_{\lambda} \,  \hbar \lambda^{-1} \, .
\label{Egraviton}
\end{equation}
The reason why the total gravitational  energy is extremely well approximated by a  simple sum of 
the energies of the individual quanta is the following. 
First,  the peak of the distribution is at  $\lambda \, =  \, R$. The contribution from the shorter wave-lengths is exponentially-suppressed  and can be ignored. 
Thus, for $R \, \gg \, L_P$, the gravitons contributing to the energy are of very long wave-lengths and thus are weakly interacting. 
 Thus, for the purpose of our estimate the interaction between the gravitons can be ignored. 
 Notice, that for $R \, \gg \, r_g$ not only the interactions between the individual gravitons can be 
 ignored, but also the interaction between any individual graviton and the entire collective gravitational energy.  In other words for $R \, \gg \, r_g$ we can ignore gravitational self-sourcing.
 This is why in this regime the condensate cannot be self-sustained.  

 Then we easily obtain  the occupation number of  gravitons by dividing the total gravitational energy by the characteristic energy of a single quantum, $N \, \sim N_{R} \sim  E_g /(\hbar R^{-1})$,  which gives us (\ref {N}).

For the future,  it is very important to note,  that even when $R \sim r_g$ the interactions among the individual gravitons  continues to be negligible. However,  the gravitational energy becomes 
of the order of the energy of the source.  At this point the self-sourcing by the collective gravitational 
energy becomes important and the condensate becomes self-sustained. 
However,  interactions among the individual gravitons is still negligible as long as  $r_g \, \gg \, L_P$.  So the occupation number of gravitons can still be safely estimated as given by (\ref{N}). 

 Since by default the physical size of the source cannot become less than $r_g$, the 
occupation number of gravitons for arbitrary source is thus universally given by (\ref{N}).  
      
  The criterion of classicality then is 
  \begin{equation}
  N \gg 1 \, .
  \label{measureN}
  \end{equation}
  The above criterion has a clear  physical meaning. 
 A given  configuration is classical when there are many gravitons in it.    We can rewrite $N$ in the  following equivalent forms, 
  \begin{equation}
   N \, = \,   {L_P^2  \over L_C^2} \, = \,  {M^2  \over  M_P^2} \, = \, {r_g^2 \over L_P^2}.
    \label{alsoN}
    \end{equation}
  The quantity $N$ is  "super-classical", in the sense that it diverges for $\hbar \rightarrow 0$.      
 This  is just a reflection of the fact that in the classical limit the number of quanta in any field configuration  is infinite.    
 
    Noticing,  that according to (\ref{Eg}) for a black hole  $M$ is equal to gravitational energy, 
 and expressing the latter via equation (\ref{Egraviton}) through $N$ and the graviton wave-length, we arrive to the following expression for $N$ in the black hole case,  
 \begin{equation}
   N \, = \, { \lambda^2  \over  L_P^2 } \, \equiv \,  \alpha_{gr}^{-1} \, .   
  \label{nforbh}
 \end{equation}
  Thus, the occupation number of gravitons in the black hole is given by the inverse gravitational coupling constant.

  The fact that $N$ is a good measure of classicality is already apparent  from the fact that for any elementary particle lighter than  $M_P$, it is less than one. 
  For example, for an electron,  $N = (m_e^2 /M_P^2 ) \sim 10^{-44}$! This is why an electron can never be regarded as a classical gravitating object despite the fact that it does 
  create a Newtonian gravitational field.  The  Newtonian gravity produced by a non-relativistic electron  does not contain even a single quantum of graviton. 
 
   Alternatively,  any source for which $N \gg 1$ 
  is classical with a good approximation.  In particular,  probability for such sources to  decay into two (or few) particle states,  are suppressed by $exp(-N)$.  
  
   By   applying  our measure of classicality we conclude that  among all possible objects of the fixed size $R$, the black hole is the most classical one,   as it contains the maximal possible number of quanta per given size. 
     Any attempt of further increasing $N$ will lead to the increase of the black hole size, 
     and correspondingly of the particle wave-lengths as $\sqrt{N}$.   This fundamental fact 
 enables  us to reformulate the black hole dynamics in terms of $N$, without involving the classical 
 geometric notions.

  \section{Universality of $N$}

  An important property of  graviton occupation number $N$  is that it is insensitive to the physical
  composition 
  of the source, but only to its mass  (or center of mass energy).  
 Obviously, with increasing 
$R$ (but keeping the mass fixed), the gravitational self-energy of the source  decreases as 
$Mr_g/R$.   So the wavelengths of the constituent gravitons increase as $R$,  and 
their effect on the dynamics of the source becomes unimportant. 
 However,  for any source that is probing distances comparable to its gravitational radius $r_g$, the $N$ gravitons become the dominant contributors to the energy and they settle the entire dynamics.
 
   The universality of $N$ represents the quantum foundation for the  universal properties of black holes (such as no-hair, thermality and entropy)  that are well known in the  (semi)classical description. 
    The same universality is also the key  for the quantum description of the phenomenon of  classicalization of gravity  in  trans-Planckian scattering of quantum particles.   
 
 The central point is that,  whenever a source is trying to probe distances comparable to its gravitational radius,  it becomes a  $N$-particle state.   This simple fact plays the key role in understanding  deep-UV properties of gravity,  since it tells  us that in gravity all the states with trans-planckian center of mass energy   $\sqrt{s} \, = \, M\, \gg \, M_P$, are in reality $N$-particle states with $N$  given  by (\ref{N}) and being equal to the entropy of an equal mass black hole. 
  
   {\it    To put it  short,   trans-planckian gravity is large-$N$ gravity. }       
 
  
  The reason why gravity classicalizes in deep-UV is due to the universality of $N$, 
  which is independent of the nature of the source.  
  The composition of the source plays no role in determining the occupation number of gravitons, 
  but only its total center of mass energy.  $N$ is the same  both for (semi)classical sources composed 
  out of large number of  relatively long wave-length particles, $N_{source} \, \gg \, 1$, as well as for quantum sources composed of few  short wavelength quanta,  $N_{source} \sim 1$.  
  
For example,  the source  can represent  an initial state of  an $S$-wave scattering of two particles 
with very high center of mass energy $M$.  Despite this, the entire state cannot be 
treated as a two-particle state and instead is a many-particle state composed out of 
 $N_{source} \, = \, 2$ source particles plus $N$ soft graviton.  The number of gravitons  does not change even 
 in $R\rightarrow  \infty$ limit.   In this limit the gravitons become infinitely soft 
but their number remains  $N$.  

 In the same process,  the similar conclusion does not apply to the occupation number of quanta of other weakly-coupled bosonic fields that are not sourced by the energy. For example, even if the particles in question are charged (say we consider a two-electron scattering), the occupation number of photons  in the initial state is of the order of the fine structure constant.,  $N_{\gamma} \, \sim \, \alpha_{EM}$. 
 This is why the  elementary charges do not represent classical states from the point of view 
 of the electromagnetic field created by them, since the occupation number of  photons 
 in this field is less than one.

     
  
Thus, any trans-Planckian source in gravity that probes distances comparable to its $r_g$-scale 
is a multi-particle state regardless of its composition. We shall discuss this in more details now.

   Until now we were  treating our source as an external classical source. In quantum field theory there are no such sources. All the gravitating sources are composed of particles. 
   We now wish to view the source from the point of view of quantum field theory and to identify the maximal number of quanta that could contribute to the composition of  such a source. 
   We shall assume that  the  underlying theory responsible for the internal dynamics of the source is in a weakly-coupled 
   regime with respect to the quanta of interest.   This means that the contact interactions 
 among particles  must be weak, but  the collective effects can be strong enough 
 to form a bound-state.
  The maximal number of weakly-coupled quanta composing the source can be estimated 
  as follows.  Since the size of the source is $R$,  due to Heisenberg's uncertainly principle, the
  minimal energy of any quantum localized within the source is $E  =  \hbar / R$.   
  Thus, the maximal number of quanta making up a source of mass $M$ and size $R$ is 
  \begin{equation}
  N_{source-max} \, = \,  MR/\hbar \, .
  \label{Nsource}
  \end{equation}
  Actually the source may not be made out of the maximal number of soft quanta,
  but rather out of  few hard ones.   As an example, consider  a pair of  colliding particles with huge center of mass energy. Thus,  the real number of quanta composing the source,  which we shall denote by $N_{source}$, may be anything  equal or below the $N_{source-max}$. 
  
    Comparing (\ref{Nsource}) with (\ref{N}), we observe the following. 
   As long as $R \, \gg \, r_g$, the number of quanta making up a source, $N_{source}$,  can in principle dominate over 
  the occupation number of the longitudinal gravitons  $N$ in the gravitational  
field produced by the source.  However,  for $R \sim r_g$ the number of gravitons always
 reaches maximal available number, irrespective of the origin of the source. 
 
  Put differently,  regardless whether in the absence of gravity the source would be quantum or classical,  with gravity it always  {\it classicalizes}  as long as $M \, \gg \, M_P$, and $R\sim r_g$. 
At this point,  any source becomes  a gravitationally self-sustained  $N$-particle bound-state, or to be more precise a Bose-condensate of $N$ soft weakly-interacting gravitons.   
 All the known properties of black holes emerge from the quantum-mechanical effects of this bound-state.  Their  level of classicality is  controlled by $N$.  
 
  Universality of $N$ enables us to abandon usual  classical and geometric notions  and formulate the quantum theory  of a black hole entirely in terms of  $N$ that controls all the essential characteristics, such as, particle occupation number, their interactions and the wave-lengths.   . 
 We shall now discuss various aspects of this  emergent black-hole picture.

\section{Black Holes's Quantum Portrait}
 
   In our picture the following simple quantum portrait of the black hole emerges which does not rely  on any classical geometric characteristics, such as horizon.  Instead,  it  is fully characterized 
 by a single quantum parameter $N$.    
  
  For us the black hole is a bound-state  (Bose-condensate) of  $N$ weakly-interacting gravitons of characteristic wave-length, 
  \begin{equation}
    \lambda \, = \,  \sqrt{N} L_P \, . 
    \label{lambda1} 
    \end{equation}
  The quantum interaction strength between individual gravitons is weak and is given by 
  \begin{equation}
    \alpha \, = \, {1 \over N} \, .
    \label{alpha1}
   \end{equation}
   Correspondingly the mass of the bound-state approximately is given by the 
  sum of  energies of individual quanta,   
   \begin{equation}
  M \, = \,  \sqrt{N}  {\hbar \over L_p}  \, . 
  \label{mass1}
  \end{equation}
 Notice,  that in large  $N$ limit, in which the geometric picture is a good approximation, 
 the wavelength $\lambda$ can be used as the characteristic size of a black hole, and the 
 horizon area  scales as $\lambda^2 \, = \, N \, L_P^2  $. This creates an impression that the horizon  represents a
 collection of $N$ cells of  Planck size area.  Correspondingly the expression for 
 the mass (\ref{mass1}) creates an impression that the mass of a black hole 
 is composed out of $N$ Planck wave-length  quanta. 
 Thus, in the geometric limit one may conclude that  these scaling laws indicate 
that black hole horizon could  secretly  represent a  probe of Planck-scale physics.  
 
  In reality  this is just an  "optical  illusion" as it is clearly indicated by equations (\ref{lambda1}) 
  and (\ref{alpha1}).   For large $N$, the black hole is a quantum bound-state  of very soft  and very weakly-coupled quanta, and by no means it represents any better probe of Planck scale physics
  than other macroscopic objects.   
  
    This quantum bound-state, however,  has some very special properties which makes it different 
    from the other similar quantum bound-states  of soft particles, such as solitons.  
 
  The defining properties are: 
  
  $~~~~$
  \begin{itemize}
 \item   {\bf 1.}  The bound-state exists for arbitrarily large  $N$.  
  
  \item  {\bf 2.}    $N$ always saturates the maximal occupation number for given wave-length ($\sqrt{N}\, L_P$).

$~~~$ Due to this,  the bound-state for any $N$ is  {\it leaky}.           
\end{itemize}      
     
   $~~~$
   
  The above statements follow from the following simple counting. 
   First note that the gravitational interaction strength of two gravitons given by 
   $1/N$ is achieved for the wavelength (\ref{lambda1}).   
  For  gravitons with  interaction strength $1/N$, the maximal occupation number for which the 
  collective interaction becomes strong is obviously $N$.   Thus, each graviton only sees 
  the potential  of the wavelength range.  Such potential is leaky, since the escape 
  energy is just above the energy of the condensed quanta.

  Alternatively we can start with arbitrary $N$ and $\lambda$ and get convinced  that the gravitational bound-state is achieved  exactly for the values  given by (\ref{lambda1}) and (\ref{alpha1}). 
    
  
     Thus, consider  $N \, \gg \, 1$  quanta of wave-length $\lambda\, \gg \, L_P$
     interacting gravitationally. 
   For large wave-lengths  the interaction strength between a pair of individual gravitons is  extremely weak  and  is given by 
\begin{equation}
   \hbar \,  \alpha_{gr} \, = \,  \hbar^2 G_N /\lambda^2 \, ,   
   \label{weakness}
   \end{equation}
 so that effectively each graviton sees a collective binding potential created by all  $N$ gravitons,  
 \begin{equation}
 V(r)\, |_{r \gtrsim \lambda } \,  = \,   \hbar  \alpha_{gr} N  {1 \over r} \, .  
 \label{potential}
 \end{equation}
  In this regime the wave function of a black hole  can be very well approximated by a Hartree-type wave function, 
\begin{equation}
\Psi = \prod_{i}^{N}\psi_i \, , 
\label{BHHartree}
\end{equation}
with $\psi_i$ being a  one-particle wave function solving the Schr\"odinger equation for the average potential (\ref{potential}) created by $N$ gravitons.
 For any fixed $\lambda$, this potential reaches a maximal depth around $r \, = \, \lambda$  and the escape kinetic energy of a probe graviton moving in such a potential  is, 
 \begin{equation}
  E_{escape} \, \equiv \, {\hbar \over  \lambda_{escape}}  \, = \,  \hbar  \alpha_{gr}  \, 
   N \,  {1 \over \lambda}  \, .
 \label{escape}
 \end{equation}
 Thus, the escape wave-length saturates the wavelength of the condensed quanta 
 precisely when the quantities  $\lambda$,  $\alpha_{gr}$, and $M$ satisfy the relations 
 (\ref{lambda1}), (\ref{alpha1})  and (\ref{mass1}) respectively.

 Thus,  the  $N$ gravitons of wavelength $\lambda \, = \, \sqrt{N} L_P$ form a self-imposed 
 bound-state. This bound-state exist for arbitrarily large  $N$, and for any $N$ it is leaky. 
 This leakage is the key reason for the emergence of the thermal properties of  the Hawking radiation.
 


 
 

  \section{Quantum Origin of Hawking Thermality}  
  
     As explained above, in our picture a black hole is an intrinsically-quantum object that represents  a cold Bose-condensate  of $N$-gravitons, of characteristic wave-length 
     $\sqrt{N} L_P$ and interaction $1/N$.    Exact classicality is only recovered in 
     $N \rightarrow \infty$ limit. 
 How does the notion of the Hawking temperature and radiation emerges in  this cold 
 Bose-condensate? 
 
  As we shall now show,  this happens through the quantum  effect analogous of 
quantum depletion of the Bose-condensate.  This is a well known phenomenon in quantum physics, which manifests itself  in the fact that in a Bose-condensate of interacting bosons even at zero temperature  there are always some particles with energies above the ground state. 
 The specifics of black hole physics is that the graviton condensate is a leaky bound-state 
 and the escape energy is just slightly above the energy of the condensed quanta. 
 Thus the system continuously  produces gravitons above the condensation energy, which 
 escape leading to the decrease of $N$.  Since the only characteristic of the black hole is $N$, the resulting condensate is another black hole with $N-1$ gravitons and the process 
 of depletion carries over.   Since the wavelength of radiated quanta  for each $N$ is $\sqrt{N}L_P$, the resulting spectrum for large $N$ reproduces the Hawking thermal spectrum, up to 
 $1/N$ corrections.  

     To establish the evaporation law,   quanta escape whenever their energy  exceeds the binding  energy $E_{escape} \, = \, \hbar/(\sqrt{N} L_P)$.   This happens as a result of a quantum process in which one of the quanta gets an effective  above-threshold energy by scattering about the background  gravitational potential.  
  
  A most probable process is a $2 \rightarrow 2$ scattering of two constituent gravitons in which 
  one of them gains above threshold energy and escapes. 
  Since the escape wavelength is $\lambda_{escape} \, = \, \sqrt{N} L_P$, both initial and final energies are small and momentum transfer is  also small.   The rate for such a process 
  to the leading order in $N$ is, 
  \begin{equation}
  \Gamma \,  =  \,  {1 \over N^2}  \, N^2 \, {\hbar \over \sqrt{N} L_P} \, .  
  \label{emissionrate}
  \end{equation} 
 The first $1/N^2$ factor here comes from the interaction strength, the second $N^2$  
 factor is a combinatoric one and $\hbar /(\sqrt{N}L_P)$ factor comes from the characteristic 
 energy of the process.  
 Thus,  viewed as a quantum resonance,  the black hole has a width  $\Gamma$.  
 
 This rate sets a characteristic time-scale  $\Delta t \, = \, \hbar \,  \Gamma^{-1}$ during which a graviton of wave-length  $\lambda_{esc}$ is emitted from the black hole. The resulting decrease of the   bound-state mass is $\Delta M \, = \, - \,  \hbar/ \lambda_{esc}$. 
 Thus, we have for the emission  rate, 
 \begin{equation}
    {d M \over dt} \, = \, -\,  { \hbar  \over  \lambda_{esc}} {\Gamma \over \hbar}  \, = \, -\,  { \hbar  \over N L_P^2}  \, . 
    \label{Npower}
    \end{equation}  
  Using the relation between the mass of the bound-state and  $N$,  
  $M \, = \, \sqrt{N} \hbar L_P^{-1}$, we can rewrite the quantum law of black hole evaporation in 
  extremely simple  form, 
  \begin{equation}
   {dN \over dt} \,  =  -  { 1 \over \sqrt{N} L_P} \, .
   \label{thelaw}
 \end{equation} 
 This equation makes a clear physical sense, since it tells us that during the time 
 $\Delta t \, = \, \sqrt{N} L_P$ the condensate emits one quantum and thus $N$ decreases 
 by one.   The half life time of the condensate is  given by,  
 \begin{equation}
  \tau \, = \, N^{3/2} L_P \, .
 \label{timelaw}
\end{equation}
 It is obvious that for large $N$ the above equations correctly reproduces the thermal evaporation 
of black holes, but without any reference to geometric concepts. 
 Indeed, defining temperature as, 
 \begin{equation}
   T \, \equiv \,   {\hbar \over  \sqrt{N} L_P} \, ,  
   \label{temperature}  
   \end{equation} 
 the equations (\ref{thelaw}) and (\ref{timelaw})  immediately translate into the familiar 
 Hawking expressions for thermal evaporation rate, 
  \begin{equation}
     {dM \over dt} \, = \, -  \, {T^2 \over \hbar}  \, , 
     \label{rateold}
 \end{equation}
  and the half life-time, 
   \begin{equation}
    \tau \, = \, {\hbar^2  \over  T^3  G_N} \, . 
    \label{timeold} 
  \end{equation}
  
  Notice, that a negative heat-capacity which is one of the most mysterious properties of black hole thermodynamics, in our quantum description is a trivial consequence  of the fact that 
$N$ has to decrease as a result of quantum depletion.

   To conclude, we have seen that Hawking temperature and radiation emerges in our picture 
  entirely as a quantum-mechanical effect analogous of quantum depletion of a leaky 
  condensate of $N$ soft weakly coupled gravitons. In large $N$ limit this reproduces the familiar 
  Hawking result, but at no point we rely on (semi) classical notions such as horizon, or temperature. 
  The classicality  for us is an emergent phenomenon, corresponding to the limit of infinite $N$ \footnote{In order to make the connection with a Bose Einstein condensate \cite{Bose} of soft gravitons more transparent we can use the Gross-Piataevskii approximation with a density $n = \frac{1}{\sqrt{N}L_P^3}$ and scattering length
 $a = \frac{L_P}{\sqrt{N}}$. This leads to a {\it healing length} $\xi = (n.a)^{-\frac{1}{2}} = \sqrt{N}L_P = \frac{1}{T}$. Note that the speed of sound in this gravitational Bose Einstein condensate is the speed of light. }.

  \section{Universality of Radiation and  Bound on the Number of Particle Species}

   One of the important  by-products  of our description is that it automatically  explains  
 both the universality of the Hawking radiation as well as the black hole bound on the number of 
 elementary particle species \cite{species}. 
   Let us assume that in the theory the number of light particle species is  $N_{species}$. 
    This number counts all the propagating  helicities, including the graviton itself. 
 Without the loss of generality, we can assume these species to be massless. 
 It is super-important not to confuse $N_{species}$ with $N$.  The former represents  a fixed parameter  of  the theory, whereas $N$ is just an occupation number of gravitons in a particular
Bose-condensate  and  assumes  different values  for different black holes. 

  In \cite{species} it was shown  that  in the semi-classical treatment  $N_{species}$ sets  an universal lower bound on the size of a smallest possible semi-classical black hole, $L_N$,  as, 
  \begin{equation}
  L_N \, = \, \sqrt{N_{species}} \, L_P \, . 
  \label{speciesbound}
  \end{equation} 
  From the first glance, appearance of this bound looks pretty mysterious. 
  How could the existence of massless non-gravitational species  dictate the theory down to which scales  black holes can exist? 
     In the semi-classical treatment the above bound appears as a consistency requirement, but 
 has no underlying explanation. 
   However,  our quantum picture provides such an explanation is very simple terms. 
 The bound on species scale $L_N$ translates as the bound  on the wave-length beyond which 
 the depletion becomes so strong that the condensate can no longer afford to be  self-sustained. 
   
    Indeed, let us again repeat the previous analysis with the depletion of the graviton  
    condensate, but now taking into  the account the existence of the extra particle species. 
   This opens-up the additional channels for the depletion.   The two constituent gravitons instead 
   of re-scattering into two gravitons, now can also annihilate into a pair of  any of the remaining 
   $N_{species}$ particle species.     Just like in the graviton case, one member  of the produced pair acquires  an escape  energy and leaves the condensate, whereas the other member 
   remains in the bound-state.  In case when the particle in question carries a conserved 
   gauge quantum number,  the resulting condensate will consist of $N-2$-gravitons and 
   an extra particle.  
    
       Due to the universality of the graviton coupling, the rate of the above  process is the same as 
       the rate of depletion in the graviton emission channel  (\ref{emissionrate}). 
    So,  the resulting total depletion rate is enhanced by the factor $N_{species}$, 
   \begin{equation}
   \Gamma \, = \, {\hbar \over \sqrt{N} \, L_P} \, N_{species} \, .
   \label{rateNsoecies}
   \end{equation} 
    Due to the universality of the process, each of the $N_{species}$ species is produced with  the same thermal spectrum
    of temperature (\ref{temperature}). The emerging  impression is that the particles are radiated 
   from an universal thermal bath, but in reality the condensate is cold. 
  
   It is now totally obvious that for $N \,  < \, N_{species}$  the rate exceeds the mass of the black hole, 
   \begin{equation}
    M\, < \, \Gamma  \, ,      
    \label{violation}
    \end{equation}
   and the condensate can no longer be self-sustained.   Beyond this point  the condensate 
   becomes so leaky that it is no longer even a bound-state: Its  width becomes broader than its mass!
   Requirement that the bound-state is self-sustained puts the bound on the occupation number 
   $N \, > \, N_{species}$, which translates as the lower bound 
  on the wave-length of the condensed gravitons as, 
  \begin{equation}
    \lambda \, > \, L_{N} \, , 
   \label{lambdaln}
   \end{equation}
  which taking into the account  (\ref{lambda1}),  exactly reproduces the bound (\ref{speciesbound}).   
  
   Thus, the  underlying quantum meaning  of the black hole bound on species (\ref{speciesbound}) is the following.   The self-sustained Bose-condensate of the gravitons can exist as long as 
   the occupation number of  condensed gravitons exceeds the total number of elementary 
  particle species in the theory.   Beyond this point the condensate depletes so rapidly that 
  it is no longer a well-defined bound-state.  No semi-classical black hole
  exists  beyond this point.

   \section{Quantum $N$-Portrait of  Solitons}

   Our quantum picture  of a black hole exhibits similarities  with the  properties of solitons.   
     Just like black holes, solitons too are objects whose entire energy is composed out of the 
energy of many  soft quanta. 
 We shall now show that this similarity is fundamental, and that quantum portraits of black holes 
 and solitons  formulated in terms of particle occupation number are identical. 
 This  fact allows us to give a unified description 
  of these objects in terms of large-$N$ quantum Bose-condensate.
  Moreover, we shall understand fundamental differences, such as, why solitons 
  do not  exhibit emergent  characteristics such as thermality or entropy, whereas black holes do.   

 We shall now  formulate an analog  $N$-particle theory of solitons on an explicit example of  
 't Hooft-Polyakov magnetic monopole. 
     These solitons exist in gauge theories with topologically non-trivial vacuum structure, 
    the simplest example being a $SO(3)$ gauge theory Higgsed down to 
    $U(1)$ subgroup.   Classically, this theory has two parameters. 
    The gauge coupling constant $g$ and the vacuum expectation value of the Higgs field, $v$. 
    The dimensionalities of these quantities are, 
    \begin{equation}
        [g] \, = \, ([mass][length])^{-{1 \over 2}}  \, ,~~~~~~   [v] \, = \, {[mass]^{{1\over 2}} \over  [length]^{{1\over 2}}}  \,  .
    \label{gandv}
    \end{equation}
    In quantum theory, forming appropriate combinations with $\hbar$, we can form a 
    dimensionless gauge coupling strength, 
    \begin{equation}
     \alpha_{gauge} \, \equiv \, \hbar \, g^2 \, ,  
     \label{alphagauge}
     \end{equation} 
    which is the analog of   $\alpha_{gr}$, and a length-scale 
    \begin{equation}
    L_{v} \, \equiv \,  \sqrt{\hbar} / v \, , 
     \label{lv}
     \end{equation} 
    which is an analog of the Planck length.     
    Let us also define the mass of the gauge boson (the $W$-boson)  $m_W \, = \, gv\hbar$.  
   Throughout our consideration  we shall stay in the weak coupling regime, 
   $\alpha_{gauge} \, \ll \, 1$.

      The magnetic monopole  represents a classical (topologically-stable) soliton of mass 
    \begin{equation}
    M \, = \, {v \over g} \, = \, { m_W \over \alpha_{gauge}} \, , 
    \label{monopolemass}
    \end{equation}   
    The physical size of the monopole is given by the Compton wave-length of the $W$-boson,  
    \begin{equation}
      R \, = \, \, {1 \over gv} \, = \, {\hbar \over m_W}  \, \equiv \, \lambda_W\, ,   
     \label{monopolesize}
     \end{equation}
    Notice, that since the mass of the $W$-boson contains $\hbar$, the size of the monopole is classical.  The reason why monopole is a classical object is because it can be viewed as 
   a  superposition of many $W$-bosons, of wave-length $R$ and energy $m_W$. Their occupation number is given by 
   \begin{equation}
    N_{W} \, = \,  { 1\over \alpha_{gauge}} \, = \, MR/\hbar. 
    \label{NW}
    \end{equation} 
   Because the theory is weakly coupled, the number of $W$ bosons saturates the maximal number 
   $N_{source-max}$ permitted by Heisenberg's uncertainty principle.   That is, the energy 
   of the magnetic monopole is approximately saturated by maximally soft $W$-bosons, 
     \begin{equation}
    M \, = \,  m_W \,  N_W \, .  
    \label{numbermass}
    \end{equation}
    
  Observe, that,  according to above equations all the characteristics of the monopole can be expressed in terms of 
  $L_v$ and $N_W$ in the following way,  
  \begin{equation}
  \lambda_W \, = \, \sqrt{N_W}\,  L_v  \, ,~~~ \alpha_{gauge} \, = \, {1\over N_W} \, , ~~~
   M \, = \, \sqrt{N_W}  {\hbar \over L_v } \,, .
   \label{analog}
   \end{equation} 
   
 These are identical to the black hole ones  (\ref{lambda1}), (\ref{alpha1}) and (\ref{mass1}). 
  We discover that the soliton and black hole quantum portraits formulated in terms 
  of occupation number are identical.   The monopole represents a bound-state of soft 
  bosons of wavelength $\sqrt{N_W}$ coupled with a strength $1/N_W$.  
  
   Just as for black holes, for large $N_W$ we can define a geometric notion of the surface area
  which contains $N_W$  elementary cells of fundamental area $L_v^{2}$,  in the 
  same way as the black hole horizon contains $N$ Planck area cells. 
    Neither black holes nor monopoles represent the probes of the fundamental length, 
  instead they probe physics at the scales $\sqrt{N}$ times  longer.   
  
  The  fundamental difference between black holes and solitons is,  that black holes exist for arbitrary 
    $N$ whereas for solitons $N_W$ is fixed. The reason is the  very different  energy  dependence 
    of the coupling $\alpha$ in the two cases.   In gravity, due to energy self-sourcing  $\alpha$ 
    depends on the graviton wave-length quadratically as given by (\ref{alphag}). Correspondingly, 
    the bound-state  exists for  arbitrary $N$.   This is why the black holes are leaky.
    Decreasing  $N$ just takes the system to a lower mass bound-state with 
    smaller $N$ and thus shorter wave-length. 
    For solitons $\alpha$ is fixed, and the decrease of $N$ is impossible.  This is why solitons 
    are not leaky.  
     Macroscopically, in case of monopole, non-depletion can be understood in terms of 
     a conserved magnetic (or topological) charge, but  microscopically this is 
     explained by  the fact that $N$ is fixed in the lowest energy state.

      Nevertheless, if we couple monopole to gravity, it will act as a source, and this will create 
   occupation number of gravitons. Now the system will consists of 
   $N_W$ gauge and $N$ gravitational quanta.     
      
    The occupation number of longitudinal gravitons in the gravitational field 
   created by the magnetic monopole is given by (\ref{N}). For  $\lambda_W\, \gg \, r_g$,   we 
   have  $N_W \, \gg \,  N$.  Thus, as long as the size of the monopole is much larger than its 
   Schwarzschild radius, the particle  occupation number  is dominated by gauge bosons and not by gravitons.   The situation changes dramatically  whenever  we try to deform monopole into a black hole. 
   
    This is easy to understand, since the monopole becomes a black hole when its size 
   crosses over within its Schwarzschild radius.  At the crossover, $r_g \, = \, \lambda_W$,  we can rewrite 
   (\ref{NW}) as 
      \begin{equation}
    N_{W} \, = \,   { m_W\over g^2 \hbar} {1 \over m_W} \, =\, 
    {M r_g \over \hbar} \, 
    \label{NWN}
    \end{equation} 
 which implies, 
      \begin{equation}
    N_{W} \, = \,  N \, .
    \label{NN}
    \end{equation} 
 The above crossover phenomenon of the occupation numbers  explicitly demonstrates that even for classical sources  (including maximal occupation number of the soft quanta),  the configuration starts to be dominated by the number of  gravitons 
 (\ref{N})  whenever the size of the source approaches its gravitational radius.  We shall 
 now  follow more closely the physics of this crossover.

  
  
  

         \section{Soliton/Black Hole Correspondence}
    As discussed above, a      
t'Hooft-Polyakov monopole can be continuously deformed into a black hole by simply changing the strength of the coupling $g^2$ while keeping fixed the value of $M_W$. The black hole threshold is determined by the condition $N=N_W$. According to (\ref{NWN}),  the critical value of the coupling 
 for which this is achieved is determined from the condition,  
\begin{equation}
\hbar g_c^2 = \frac{M_W^2}{M_P^2} \,  = \, {L_P^2 \over \lambda_W^2}
\label{gcrit}
\end{equation} 
For $1>>g>g_c$ the monopole can be thought as being composite of  large number $N_W$ of weakly-interacting soft quanta with coupling constant $\hbar g^2=\frac{1}{N_W}$. In this regime the wave function of the monopole can be nicely approximated by a Hartree wave function 
\begin{equation}
\Psi = \prod_{i}^{N_W}\phi_i
\end{equation}
with $\phi_i$ the one particle wave function solving the Schrodinger equation for the average potential created by $N_W$ particles. In the particular case of the monopole the average potential acting on one particle states includes the binding magnetic field of the monopole.
 Note that the strength of this magnetic field increases with $N_W$. 
 So,  for $g \, \gg \,  g_c$  the  $W$-boson condensate is self-sustained, but the graviton 
 condensate is not.    This is because  there are too few gravitons for creating enough 
 binding potential for a given wave-length.  So, the graviton condensate exists as long as 
 it is sourced by the gauge boson one, which is self-sustained. 

By gradually  reducing the gauge coupling $g$ we 
reach the critical value $g_c$ at which the number of gravitons $N$  catches up with $N_W$.  
At this point graviton condensate also becomes self-sustained. 
 So the monopole becomes effectively composed out of the equal numbers   $N = N_W$ of soft gravitons and  gauge bosons.  Both gravitational and gauge  interaction strengths  at this point are  equal  to  $\frac{1}{N}$. 
 Notice that  according to (\ref{gcrit}),   for   $g=g_c$ the magnetic charge of the monopole becomes equal to its mass in Planck mass units, 
 \begin{equation}
 Q_{magnetic} \, = \,  {1 \over g_c}  \, = \, \sqrt{\hbar} {M \over M_P} \, . 
 \label{extremal}
\end{equation}
 Thus the state $N=N_W$  coincides with the extremal magnetically-charged black hole. 
 Hence, from a quantum point of view an extremal black hole is a state in which occupation numbers 
 of gravitons and the gauge fields are equal. 
 
  The quantum picture described above explains all the well-known geometric properties  of magnetically charged black holes (see \cite{magneticbh})  in simple quantum-mechanical terms.  For example,  the classical condition of no-naked singularity, 
  \begin{equation}
  Q_{magnetic} \, \leqslant  \, \sqrt{\hbar} {M \over M_P}  \, , 
  \label{nonsingular}
  \end{equation}
which is pretty mysterious in classical language, in our picture has a simple physical meaning and 
implies, 
\begin{equation}
N_W \,  \leqslant  \, N \, .
\label{nonsingularN}
\end{equation}
As we have shown, violating the above condition pushes us back into the no-black hole domain $N < N_W $ and simply describes a gravitating 
magnetic monopole fully dominated by $N_W$ gauge quanta of wave-length $\hbar /M_W$.  
Ignoring this underlying quantum consistency and seeking for a 
solution in form of  Reissner-Nordstr\"om black hole  without 
massive  $W$-boson fields outside the horizon, one is "punished" by the appearance of the 
naked singularity!
 In other words, classical naked singularity is the way the theory is trying to tell us that we are dealing with an unphysical solution  that does not satisfy the underlying quantum mechanical 
 consistency requirements.

 Finally,  for $g<<g_c$ we have  $N \, \gg \, N_W$ and in this regime 
  the dominant dynamics is governed by the gravitational interaction among the $N$ soft gravitons. In this regime the leading contribution to the Hartree wave function comes from the average gravitational potential. As we have described in the previous section,  in this regime the wave-length of the constituent gravitons is about the escape wave-length and consequently  quantum 
 depletion of the black hole will take place.   Semi-classically, this process is seen as  Hawking radiation. 


\section{Baryon/Black Hole Correspondance}

We can think of a different example of soliton /black hole correspondence using  $SU(N_C)$ QCD with large number of colors.  The role of solitons in this case are played by baryons 
 which can be interpreted as solitons composed out of $N_C$ quanta (quarks/pions) \cite{Witten}. 
These quanta are weakly-interacting,  with the strength $\alpha_{QCD} \, \equiv \, \hbar g^2=\frac{1}{N_C}$. The characteristic wave-length of the constituent quanta, which defines 
the size of the baryon is $\lambda_{QCD} \, \equiv \, \hbar /\Lambda_{QCD}$, where 
$\Lambda_{QCD}$ is the usual QCD scale. The mass of the baryon is 
$M\, =\, N_C \, \hbar/\lambda_{QCD}$. 
 Defining the length scale $L_{QCD} \, \equiv  \,  \lambda_{QCD} /\sqrt{ \alpha_{QCD}}$, we can rewrite all the baryon characteristics in a very simple form, 
   \begin{equation}
  \lambda_{QCD} \, = \, \sqrt{N_C}\,  L_{QCD}  \, ,~~~ \alpha_{QCD} \, = \, {1\over N_C} \, , ~~~
   M \, = \, \sqrt{N_C}  {\hbar \over L_{QCD}} \, .
   \label{analogB}
   \end{equation} 
This form is a complete analog of  the similar expressions  in the black hole  (\ref{lambda1}), (\ref{alpha1}), (\ref{mass1} ) and monopole (\ref{analog}) cases, respectively.   Thus, the
quantum  $N$-portrait of baryons is identical to the one of a black hole  in which 
the role of the Planck length is assumed by the scale $L_{QCD}$. 
 From above it is obvious that  the  baryons are not the probes of physics at $L_{QCD}$- length, but rather they probe physics at length scales that are  $\sqrt{N_C}$ times 
 longer.  This is fully analogous  to the black hole case, which are probes of physics at 
 distances $\sqrt{N}$ longer than the Planck length.  
 
  The above picture also highlights obvious  differences between the large $N_C$ baryons 
and black holes.   Just like solitons, the baryons exist only for fixed value of $N_C$ set by the number of colors.  
This is why solitons and baryons are not {\it leaky} bound-states and  do not exhibit the phenomenon of quantum depletion.  This goes in sharp contrast with black holes which exist for 
any $N$ and therefore undergo quantum depletion.

 Just like we did in the monopole case, we can couple baryon to gravity and cross it over into a black hole phase by continuously deforming the parameters of the theory. 
As we did in the case of the monopole,  we can keep fixed $\lambda_{QCD}$ and change the value of $g^2$. Geometrically, the black hole threshold is achieved when the 
gravitational radius becomes equal to the size of the baryon, $\lambda_{QCD} \, 
= \,  MG_N$. Quantum mechanically this implies  $N_{source}\, = \,  N_C \, =\, M^2 /M_P^2$. 
The critical value of the coupling  is given by
\begin{equation}
\hbar g_c^2= \frac{L_P^2}{\lambda_{QCD}^2} \, .
\end{equation}
At this point, the occupation number of gravitons becomes equal to the number of colors, 
$N \, = \, N_C$ and the graviton condensate becomes self-sustained. 

The Hartee wave function of the baryon for $1>>g>g_c$ is defined by an average potential acting on one particle states that simulates the binding force of a QCD string of tension 
$\mu_{string} \, = \, \hbar /\lambda_{QCD}^2$. At the critical value $g \, = \, g_c$ the baryon starts to be dominated by purely gravitational quanta becoming a full-fledged black hole for $g\, \ll \, g_c$. Note that at the crossover the gauge coupling agrees with the string coupling of a 
QCD string theory with the string length $L_s \, = \, \lambda_{QCD}$. In the black hole regime $g\, \ll \, g_c$,  we have $N_C \, \ll \, N$ and system "forgets" about its baryonic origin.   In this domain 
of the parameter space, the dynamics  is dominated by the graviton condensate, which can deplete according to  (\ref{rate}), as discussed above. 

In both soliton and baryon examples the gauge-gravity correspondence takes place at the crossover point $g \, = \, g_c$ (equivalently $N_{source} \, = \, N$)   where the Hilbert space of the constituent gravitons and the one of the gauge constituents of the source become isomorphic. Again it is important to stress here the universality of the gravitational Hilbert space of the $N$ soft gravitons. This space is the same irrespective of the nature of the source we start with and it is this universality the key ingredient we will need to uncover the notion of black hole entropy.

Summarizing in the previous examples we have developed a unify large $N$ portrait of solitons, baryons and black holes. The specific feature of {\it large $N$ gravity} lies in energy self sourcing. This basic fact self tunes the coupling to be $\frac{1}{N}$.

\section{Entropy} 
  
  
   We now wish to clarify the key reason underlying the interpretation of $N$ as an entropy.  In more precise terms how the universality of $N$ gives us 
  an understanding of the origin of black hole entropy in terms of occupation number of gravitons.

    As it was already noticed,  for any source the number $N$ 
    coincides with the Bekenstein entropy of the corresponding mass black hole.  This remarkable coincidence is not  accidental, and has a deep physical reason.

  
  
  The universality of $N$ naturally indicates that $N$ is a legitimate {\it quantum}  characteristic 
  of a black hole, just like entropy is in the semi-classical limit.  Is the coincidence of these who quantities accidental? 
  
  In our physical picture of a black hole as a Bose-condensate of $N$ long-wavelength weakly-interacting 
  gravitons, the notion of entropy can only  emerges as the number of  quantum states 
  in which these $N$ gravitons can exist.   We wish to show now that  this number increases 
  exponentially with  $N$, and thus coincidence between  $N$ and  entropy is not accidental. 
  
  Of course, in our analysis we shall limit ourselves by order of magnitude estimates. 
  Derivation of precise coefficients is beyond the scope of the present paper.  
  
  Our task is thus to understand in how many states the $N$ constituent gravitons could be. 
 If gravitons were non-interacting and indistinguishable, then the number of states would scale only as 
 power of $N^{\alpha}$, with $\alpha$ being determined by the number of states of a single graviton. The reason why the number of states instead is exponentially large is because gravitons 
 do interact, and  because of this interaction the wave-function of the Bose-condensate can be viewed  as a direct product of wave-functions of distinguishable  {\it flavors}.  In first approximation the total number of states 
 will then be given by the product of number of states for individual flavors, 
 \begin{equation}
 n_{states} \, = \, \prod_j \xi_j
 \label{nstates}
 \end{equation}
 where $j\, = \, 1,2, ...N_{flavor}$ labels the independent flavors and $\xi_j$ is a characteristic number of states.  Let us estimate the number of unions $N_{flavor}$. 
 
  Notice, that such different flavors can only exist, if we can think of the original Bose-condensate 
  as the bound-state of sub-bound-states. 
   Such sub-bound-states do indeed exist. 
   Recall, that the very reason why black hole exists for an arbitrary $N$ is because $N$ 
   gravitons of wave-length $\sqrt{N}L_P$ can always form a leaky bound-state of the 
   same size and of mass $\sqrt{N} \hbar/L_P$.   This is an exceptional property of gravity and  has to do with the energy self-sourcing.  Since $\alpha_{gr}$ is quadratically wave-length dependent,  
 for any number $N$ an appropriate wave-length can be adjusted  which matches the escape 
 wave-length.   Thus,  a subset of $N_j$ constituents can form an union of wave-length 
 $\sqrt{N_{\alpha}} \, L_P$ and of energy $M_{\alpha} \, = \,   \sqrt{N_{\alpha}} \hbar /L_P$.   The flavor is defined  as a set of $\alpha  \, = 1,2,...n_j$ unions that form a bound-state of the mass 
 equal to the mass of the  black hole $M \, = \, \sqrt{\sum_a N_a}$. Of course, the possibilities 
 are limited by total mass (size) and the total number $N$, which demands that in the leading order, the unions must satisfy $\sum_a \,  N_a \, = \, N$. 
  
     Then to first order in $1/N$, the wave-functions of such flavors are orthogonal and form eigenstates of the Hamiltonian  with energy equal to the black hole mass.   
  
 Because of the above mass and the total number constraint, the number of possibilities 
 is cutoff by the cases when the number of unions is of order one, 
 and correspondingly the number 
 of their constituents is of the same order as the total number,  $N_a \, \sim \, N$.
 This implies that the number of flavors $N_{flavor}$ grows as $N$.
 

 Thus, the wave-function of the black hole represents (to first order in $1/N$) a direct product of  non-interacting 
one-flavor states, 
\begin{equation}
 \Psi_{BH} \, = \, \prod_{j}^{N_{flavor}}\,  \psi_j 
\label{BHwavefunction}
\end{equation} 

 Then the total number of states is given by (\ref{nstates}).  
 In other words in each order in $1/N$ each flavor acts as a distinguishable one-particle non-interacting state of degeneracy $\xi_j$.   It is obvious that this scaling is exponential with $N$.   
  
  Indeed , since the flavors have 
  similar characteristics (they represents a soft bound-state of soft sub-bound-states, with 
  all the occupation numbers  being of order $N$), the number  
 of possible states for each union $\xi_j$ must be comparable 
and can be approximated by a common number $\xi$.  The number of states then 
scales as, 
\begin{equation}
 n_{states} \, \sim \, \prod_j^{N}  \,  \xi \, = \, \xi^{N}  \, . 
\label{unions}
\end{equation}
 The above relation shows that the resulting entropy should scale as $N$, 
 \begin{equation}
 s \, = \, {\rm log} (n_{states})  \, \sim \, N
 \label{Nentropy} 
\end{equation}
   To summarize,  the connection between $N$ and the entropy 
  amounts to the fact that the universal number $N$ dictates the number of micro-states  
of the resulting configuration. 
Because the gravitons in the Bose-condensate  are  $1/N$- coupled, they can form 
$\sim N$ distinguishable flavors  and the  number of states must scale as $\xi^N$, where $\xi$ is some characteristic number of states in which a flavor  can be.  
 The important fact is the exponential scaling with $N$.  
 
  We are now ready to understand why the solitons (or large-$N_C$ QCD baryons) 
 do not exhibit Bekenstein enropy, despite the fact that their $N$-portraits 
 are very similar  to the one of a black hole.  However, there is one crucial difference: 
  Neither in solitons nor in  baryons the interaction of the constituent quanta is based 
  on energy sourcing.   For the weakly-coupled gauge theories, $\alpha_{gauge}$ is a fixed  
  wave-length-independent parameter.   As a result, the solitons and baryons exist only for 
  the fixed values of $N$.    As we repeated several times, this is why solitons are not leaky and cannot deplete.   Consequently,  for these objects there is no emergent thermality  in the semi-classical limit. 
   Due to the same reason,  constituents of the solitons cannot form flavors and 
the degeneracy of the micro-states is negligible.    
  
   As we shall see in the next section,  due to he same lack of energy self-sourcing solitons cannot play the role in unitarizing high-energy scattering amplitudes.  The probability of soliton-production  in two-particle collisions 
 is exponentially-suppressed at arbitrarily high energies.    This is why solitons cannot  play the role  in UV-completion of the theory by classicalization.

 
    

 \section{Entropy and Self-UV Completion}
 
 The microscopic understanding of the black hole entropy is intimately related to the most general question on how to complete quantum gravity in the UV. 
 In \cite{gia-cesar} we have suggested that gravity self-completes in the UV by classicalization. This form of self-completion goes beyond the Wilsonian paradigm and, as already stressed, deeply relies  on the increase of $N$ with energy. 
  We now wish to discuss how the quantum  $N$-language 
 establishes fundamental  connection between the emergent entropy on one hand and 
 UV- self-completion by classicalization and unitarity on the other.   
  
   We shall clarify the following aspects.   The emergent macroscopic arrow of time;  Difference between black holes (classicalons) and solitons;  Connection between the entropy and unitarity. 
  We shall discuss all these in quantum large-$N$ language, not relying on any classical geometric notions. All the semi-classical concepts will emerge from quantum considerations.

\subsection{Macroscopic  Arrow of Time and Difference Between Black  Holes and Solitons}

  According to the idea of self-UV-completion by classicalization the  scattering amplitudes 
 at  trans-Planckian energies become dominated by production of the classicalized  states.
 The possibility  of black hole formation in trans-Planckian scattering has a long history \cite{transplanck,  Veneziano}, but  the idea of self-completion puts this process  in a very different light viewing black holes as the ordinary quantum states of large occupation number, $N$, 
 of soft gravitons  \cite{gia-gomez-alex}.   The self-completion by  classicalization then manifests itself  as the increase 
 of $N$ with energy.  What we are trying to stress is that it is  this increase that replaces the notion  of the 
usual Wilsonian renormalization. 
 
   This formulation of the idea creates  the two seeming puzzles: 
   
   $~~~$

    1)  From the first glance, the appearance  of the arrow of time  seems to be in conflict with unitarity.   Indeed, the  idea of UV-completion by  classicalization implies  that  in two-in-two 
    scattering classical configurations, such as black holes,  are created with high probability, whereas their decay back into two-particle 
    states is exponentially suppressed.   This creates an impression, that the inverse processes  have different rates.

    $~~~$

     2)  The naive intuition,  based on the experience with weakly-coupled gauge theories, 
tells us that production of the classical states,  such as solitons,  must be exponentially suppressed by a factor $e^{-S}$, where $S$ is the action.   So why  are the black holes different from solitons?

 $~~~$
 
 
However, the above questions  can only be puzzling as long as we stay in the semi-classical treatment, in which we ignore the quantum dynamics of the black hole constituents. 
 Let us now show how taking into the account the large-$N$ quantum  dynamics clarifies the story.


 First, the appearance of the macroscopic arrow of time has a very well defined 
microscopic quantum origin and is a direct consequence 
 of large number $n_{states}$ of  the micro-states of the Hartree wave-function (\ref{BHwavefunction}).   
 In our quantum picture the black holes (just like solitons) are  bound-states of 
 $N$ soft weakly interacting quanta.  So the high energy scattering process in which 
 the black hole (soliton)  is formed in scattering of two particles, 
 \begin{equation}
 2 \rightarrow  black~hole~or ~soliton \, ,  
 \label{process}
 \end{equation}
quantum-mechanically can be  understood as the process 
  \begin{equation}
 2 \rightarrow  N  \, .   
 \label{processN}
 \end{equation}
Of course the system is perfectly time-reversal invariant, and the amplitude of the inverse 
process 
\begin{equation}
N \,  \rightarrow \, 2 \,, 
\label{inverseN} 
\end{equation}
is exactly the same.   The  macroscopic  arrow of time emerges because the number 
of quantum $N$-particle micro-states $|N, micro\rangle$ that live in  the Hilbert  space of the Hartree wave-function (\ref{BHwavefunction}) and which 
 classically would correspond to the same  black hole
macro-state  $|N \rangle$ 
is  exponentially large.   The  probability  of formation of each of these micro-states out of the initial $2$-particle state is exponentially small, but the large number of  channels 
compensates this suppression. 

So the rate of the process $2 \rightarrow N$  in reality is
composed out of the exponentially large number ($n_{states} \, \sim \, \xi^N$)  of partial rates of different 
$N$-particle micro-states,
\begin{equation}
|\langle 2|N\rangle |^2 \, = \,  \sum_{micro-states}^{n_{states}} |\langle 2|N,  micro\rangle|^2 \, =\, 
n_{states}  \, P_{micro}  \, ,
\label{summopverstates}
\end{equation}
where $P_{micro}$ is an average probability of producing one of the micro-states. 
For the black hole  $P_{micro}$ scales as $1/n_{states}$, and is compensated 
by the large number of states.  This  is how the black holes avoid exponential suppression 
in their formation rate.   However, the solitons are behaving differently because of the lack of 
energy self-sourcing. 


In order to identify as clearly as possible what is the specific feature of the black hole,  let us start considering a generic lump of $N$ soft quanta in which the wave-lengths, couplings and the total 
mass  are given by (\ref{lambda1}), (\ref{alpha1}) and (\ref{mass1}) respectively. 
 As discussed above,  since the  quanta interact very weakly  (with a coupling of the order 
 $\alpha \, = \, \frac{1}{N}$), the wave-function of the lump can be modeled in terms of a Hartree wave function (\ref{BHwavefunction})   which is a direct product of  $N_{flavor} \sim N$  distinguishable non-interacting flavor wave-functions.  As we have seen,  such an argument  leads us to associate with this lump an  entropy of order $N \, = \, ln \, (n_{states}) $.

 Let us now imagine a quantum process in which the entire Bose-condensate decays into two particles,  or a time reversal process corresponding to the production of the condensate from 
a two-particle state.  We wish to compare the relative rates for such processes in gravity and 
gauge cases, and show that gravity gets additional enhancement  due to the fact that 
$\alpha_{gr}$ is energy-dependents. 

  Quantum-mechanically the process in which an entire Bose-condensate decays 
  into a two-particle state is described by a set of diagrams in which $N$ initial constituents 
  merge into two particles of a  center of mass energy given by $\sqrt{N}\hbar /L_P$ (the mass of the 
  original $N$-particle bound-state).  
    In order to be concrete,  let us pick up one such process in which $N$ particles cascade into 
    two via three-point vertexes.  The over-all combinatorics is unimportant since it is the same 
    for both gauge and gravity cases.  So let us consider a fixed diagram and compare 
  scalings with $N$ of the  amplitudes in the  gravity and gauge cases.   

 It is obvious that the process consists of $N-2$ elementary acts (vertexes)  in each of which the number of bosons is 
 reduced by one.   We can order the vertexes and label them with and index $j$.  
 The probability of a given act $j$  is controlled by the corresponding effective coupling $\alpha_j$ that controls the merger of  Bosons in the vertex. This is because 
 the momentum flow in the vertex changes with the change of the total number of particles. 
   We are labeling the acts, because the strengths depends on the sequence of vertexes.     
 Consequently the rate of the process  goes as 
 \begin{equation}
   \Gamma \, \propto \,   \prod_j  \alpha_j \, . 
 \label{order}
 \end{equation}
The main difference between the gravity and gauge cases comes from $j$ dependence 
of $\alpha_j$-s.   For the gauge case, the interaction strength is constant 
(we ignore logarithmic running of couplings in a three-level diagrams),   $\alpha_{gauge} 
= 1/N$,  and is independent of the 
order  of vertexes. So the rate goes as,  
 \begin{equation}
   \Gamma_{gauge}  \, \propto \,    \alpha_{gauge}^N \, = \, {1 \over N^N} \, = \, e^{-Nln(N)} \, . 
 \label{gaugeorder}
 \end{equation}
But with gravity, the story is very different, since the coupling constant $\alpha_{gr}$ 
depends on the momentum flowing through the vertex.
 Since $N$ particles cascade down to $2$, the initial energy gets  pumped into the less and less 
 particles and the momentum flowing through the vertex increases.  
 This gives a relative enhancement in the gravity case. 
 This extra  enhancement appears per  process  and is unrelated to the  overall combinatorics of the diagrams and is not removable by re- summation. 
 Physical reason for its appearance  is that gravity is self-sourced by energy.  

For example,  we can imagine a concrete  process contributing into the   decay $N\rightarrow 2$ as a cascade of the following type. We divide mentally the set of quanta into one quantum and the remaining group of $N-1$ quanta. Now,  the chosen quantum  merges with one in the group leading to one quantum of the energy  $\frac{2} {\sqrt{N}L_P}$ and a group of $N-2$ soft quanta of energy
$\frac{1}{\sqrt{N}L_P}$ per quantum. Then we repeat the process $N-1$ times  until finishing with two quanta.

 Since in every $j$-th step the momentum flowing through the vertex increases as $j$, 
 the gravitational coupling  increases in the same way, whereas in the gauge case the coupling 
 stays the same in each step. 
 
  Hence in this process the relative enhancement of the amplitude in gravity case is given by, 
 \begin{equation}
{A_{gravity} \over A_{gauge}} \,  = \,  N! \, .
\end{equation}


\subsection{Entropy from Self-Completion and Unitarity}

Until this point we have shown that for a black hole, a  self-bounded lump of soft gravitons, the  entropy should match with the decay (production) rates of  the micro-states. Let us now see how this matching is related with unitarity.

%
 An important feature of the classical gravitational radius $r_g(E)$ is to set the asymptotic behavior of the total cross section for a two-graviton scattering at a center of mass energy $s=E^2$ as
 \begin{equation}
 \sigma_{tot}(s) \, \sim \, r_g(\sqrt{s})^2 \, . 
 \label{crossr}
 \end{equation}
But,  $r_g$ is a classical notion. 
In the quantum language of $N$, this growth translates as 
 \begin{equation}
 \sigma_{tot}(s) \,  \sim  \, N(\sqrt{s}) \,  L_P^2\, . 
 \label{crossn}
 \end{equation}
 
  The quantum meaning of this growth was already explained in  \cite{gia-gomez-alex} and 
  the analysis of the present paper fully confirms that explanation.  The reason for growth 
  is the universality of $N$. As we have explained,   irrespective of the initial occupation number of the source, 
  and in particular for initial  two-particles  ($N_{source} \, = \, 2$), the scattering takes place 
  whenever the occupation number of gravitons $N$ starts dominating the energy. 
  This {\it universally} happens  for the wave-length $\sqrt{N}L_P$.  Hence the linear growth of the cross section  with $N$.  Notice that,  as we have witnessed many times in this paper, 
in this formulation of the scattering  problem there is no reference to geometry, 
  it is fully emergent from underlying quantum dynamics.  
 What is the  meaning of the equation (\ref{crossn})  from the point of view of unitarity?

 Assuming the unitarity of the S-matrix,  this behavior of the total cross section leads, through the optical theorem, to an interesting relation between the discontinuities ( in the $s$ plane ) of the $(2 \rightarrow 2)$ forward scattering amplitude and $N(\sqrt{s})$, 
 \begin{equation}
 \label{one}
 Im A(s,t=0) = N(\sqrt{s})^2 \, .
 \end{equation}
For $s$  larger than $\hbar^2 L_P^{-2}$ the main contribution to the imaginary part of $A$ comes from saturating the forward scattering amplitude by a resonance in the $s$-channel composed of $N(\sqrt{s})$ soft gravitons. The  width of this resonance
\begin{equation}
\Gamma (s) \,  \sim \,  {\hbar  \over  \sqrt{N} \, L_P} \, , 
\end{equation}
leads to the typical lifetime of the bound-state of   size $\sqrt{N}\,  L_P$. 
 This  precisely matches the rate of the black hole's  quantum depletion given by 
 (\ref{emissionrate}). 
Note that the {\it classicality} condition  $N \, \gg \, 1$  (equivalently $r_g>>L_C, L_P$)  becomes the condition of long-lived resonances $\Gamma << M$. In other words the growth of the total cross section as $N(\sqrt{s})$ reflects the classicalization of the intermediate resonance. 

From the point of view of the previous discussion it is easy to understand what is happening. The initial highly energetic state $|2;E>$ with low occupation number of hard quanta classicalizes into a state with large occupation number $N(E)$ of soft quanta with typical wave length 
$\lambda \, = \, \sqrt{N(E)} L_P$. This set of quanta form a bound state whose life time is determined by the imaginary part  of $A(s,t=0)$. In general, we shall refer to this self bounded lump as a classicalon.

The imaginary part of the forward scattering amplitude obviously depends on the amplitude of the  classicalon creation  as well as on the  time-reversal amplitude for the classicalon to decay into the two-particle initial state. Therefore,  if the typical amplitude $A(N \rightarrow 2)$ for the decay of the self-bounded lump into the initial two-particle state is as we have argued is  exponentially suppressed,  then one of the following possibilities should take place
\begin{itemize}
\item  Either the classicalon is not saturating the scattering amplitude;
\item   Or the total cross section is not behaving like $N(\sqrt{s})$;
\item   Or the classicalon carries an entropy $S$ such that $e^S |A(N \rightarrow 2)|^2 \sim O(1)$
\end{itemize}
The statement, that the black hole ( understood as the self-bounded lump of $N$ soft gravitons) is saturating (at ultra-Planckian energies ) the scattering amplitude, relies exclusively on the fact that the intermediate state is necessarily composed of $N$ soft gravitons and that this set of quanta satisfies the above-discussed conditions to create a bound-state. Moreover, 
the asymptotic behavior of the total cross section as $r_g(\sqrt{s})^2$,  which in terms of 
$N$ translates as $NL_P^2$,  simply reflects that $r_g(E)$ is the length-scale at which for arbitrary 
initial occupation number of the source $N_{source}$ the 
energy of the system starts to be dominated by $N$-gravitons.   
Therefore it follows  that the only possibility is to endow the lump of soft gravitons with an entropy that compensates the exponentially suppressed decay $A(N \rightarrow 2)$. Using now our previous estimate of $A(N \rightarrow 2)$ we conclude, in full agreement with our earlier findings  that the entropy goes like $S \, \sim \, N$. 
 
 The previous discussion allows us to identify how gravity self-completes in the UV in a 
 non-Wilsonian way via classicalization. The self-completion involves two fundamentally-connected steps.   
  One step is the automatic transformation of highly energetic states with low occupation number into classical states with large occupation number of soft gravitons. The occupation number $N$ is universal and independent of the peculiarities of the initial highly energetic state. The second step, needed for unitarization in the UV, is the  emergence of entropy (a degeneration of states of the order $e^N$) and the matching of this entropy factor and the quantum decay rates.  The point however is, that the two steps follow from each other. 
The high cross-section for the  production of $N$-particle states implies their high level of degeneracy, and vice versa,  
entropy implies that the $N$-particle states become the most probable ones, since they have the highest degeneracy of  states.

  \subsection{$t$ -- Channel View}
In the previous discussion we have addressed the meaning of $N$ as of the imaginary part of the forward scattering amplitude. It could be worth to say few words on the $t$-channel meaning of $N$. If we consider $(2 \rightarrow 2)$ graviton scattering angle in the eikonal approximation we can approximate the sum over horizontal ladder diagrams in stationary phase approximation. Interestingly enough the leading contribution allows us to uncover the number of rungs $N_{rungs}$ of the dominant ladder. This number depends on the center of mass energy and goes precisely as $N(E)$
 \begin{equation}
 N_{rungs}(E) = N(E) \, .
 \end{equation}
 This already tell us something about the physical meaning of $N$. In fact, since $N$ is the number of rungs in the $t$-channel ladder,  the effective value of $t$ is given by 
 \begin{equation}
 t_{eff} \sim \frac{t}{N} \, .
 \end{equation}
 This means that irrespectively how large and ultra-planckian could be the value of the initial $t$ the effective transfer of momentum is bounded by $\frac{\hbar}{r_g(E)}$ \cite{Veneziano},  \cite{Giddings}.  In semi-classical geometric language, this is of course the typical momentum-transfer we will expect if a classicalon of size $r_g$ has been formed.  In our language, 
 this amounts to the formation of $N$-particle bound-state of wavelengths $\sqrt{N}L_P$. 
  In other words,  we can think of the dominant $N_{rungs}$ as the constituents of the {\it classicalon resonance} in $t$-channel. This simple discussion leads us to imagine the existence of some deep connections between old fashion $s\rightarrow t$ duality and classicalization that we hope to address elsewhere.

\subsection{Holography}

 It could be instructive to compare our large-$N$ picture  
 and holography \cite{tHooftSusskind}. It is not a coincidence that the number of soft 
 four-dimensional gravitational constituents $N(E)=r_g\, E$ is precisely equal to the number of holographic degrees of freedom we can locate on the black hole horizon 
 \begin{equation}
 N_{hol} = \frac{r_g^2}{L_P^2} = N \, . 
 \end{equation}
 However the interpretation of both identical numbers is dramatically different.
 First, holography relies on a geometrical notion of the horizon screen,  which by itself is a  classical concept.  
   In the holographic case we are talking about imaginary degrees of freedom with some internal {\it information quantum number} located on a horizon screen of lower dimensionality. In this case the number of states is $e^{aN}$ for some coefficient $a$ depending on how information is encoded in terms of these lower dimensional bits. In the case of classicalization $N$ is instead the number of perfectly normal four-dimensional soft gravitons. This number determines the typical occupation number of the state we are dealing with,  making it classical when $N\, \gg \, 1$. The degeneration is not a consequence of endowing these modes with some extra internal quantum number,  but instead,  a consequence of the appearance of collective modes of the classicalized  large-$N$ state.

\section{Generalization to AdS/dS spaces?}

 Our approach of thinking about the geometry in terms of a  Bose-condensate of large 
 graviton occupation number can be generalized to other  gravitational backgrounds. 
  The maximally symmetric spaces, such as de Sitter (dS) and anti de Sitter (AdS), are 
  of special interest.  Just like black holes, classically, they are described by 
  a single parameter,  the curvature radius $R$, which replaces  the notion of the gravitational 
  radius in case of a black hole. 
  
   Of course, there are some important differences.  For instance, the AdS and dS spaces are not asymptotically 
   flat. So, the notion of the graviton occupation number can be defined within the region of the curvature radius.  Nevertheless,  we can proceed the same way as for the black hole case.  Then, translated in terms of  graviton occupation number  the $N$-portraits of  AdS and dS spaces are similar 
   to the  black holes. In a general $D$ space-time dimensions, the maximally-symmetric 
   space of a curvature radius $R$, can be thought of a quantum Bose-condensate 
  of gravitons  of wave-length $\lambda \, = \, R$ and  the
   occupation number   
 \begin{equation}
 N \, = \, {R^{D-2}  \over  L_{D}^{D-2}} \, ,   
 \label{ads}
\end{equation}
 which is nothing but a generalization of  (\ref{alsoN}) for $D$-dimensions.  
 Rewritten in terms of $N$, the wavelengths and interactions strengths satisfy 
  \begin{equation}
 \lambda\, = \, N^{{1 \over D-2}} L_{D} \, , ~~~\alpha_D \, = \, {1\over N} \, ,  
 \label{Ddimensions}
 \end{equation}
 which is again a generalization of  our equations (\ref{lambda1}) and (\ref{alpha1}) to  
 general  $D$-dimensions.  
 
  It thus appears,  that the $N$-portraits of the black holes and of maximally-symmetric spaces 
 share some similarities.  The difference is,  that just as for solitons,  for maximally 
 symmetric spaces $N$ is a fix parameter of the theory (fixed by a the value of the cosmological constant) and cannot assume an arbitrary value. 
  Implications of this picture for cosmology still has to be understood. 
 
 Finally, it would be interesting to understand what is the connection (if any) between our 
  approach and  AdS/CFT correspondence \cite{Ads}. 
  To look for a connection one has to adopt our language of thinking about geometry 
  in terms of occupation numbers of quanta.   As explained above, if we adopt this point of view 
 we can think of  AdS space as of a Bose-condensate of gravitons with 
 a characteristic wave-length of the AdS curvature radius,  $\lambda \, = \, R_{AdS}$,  and the 
  occupation number  
 \begin{equation}
 N \, = \, R_{AdS}^3 / L_{5}^3 \, ,   
 \label{ads}
\end{equation}
where  $L_{5}$ is the five-dimensional Planck length.  In terms of $N$, the wavelengths and 
coupling strength, satisfies (\ref{Ddimensions}) with  $D=5$.  
 From AdS/CFT point of view, the interesting observation is that  $N$ in (\ref{ads})
 coincides with the central charge of CFT.  
  Thus, in our language the latter charge is set by the occupation number of gravitons 
  in  AdS Bose-condensate. 
   Whether this is just a remarkable coincidence or a signal of 
  a deeper connection,  must be investigated further.  
 
 This coincidence could be telling us, that AdS/CFT correspondence could be a particular manifestation  of a much more general pattern that we are observing in this paper 
 over and over again, that physics of maximally-packed Bose-condensates gets oversimplified. 
 Whether this is the right intuition for AdS/CFT case,  remains to be seen.

\section{Final Comments: Classicality Beyond the Wilsonian Paradigm.}     
   Perhaps the most interesting output of classicality is to identify the special features of gravity in {\it geometry-independent} terms. The key point is that in a class of theories where the energy sources the interaction,  unavoidably high-energy low-occupation-number states become states with large occupation number of weakly-interacting soft constituents which  behave classically. Obviously,  we can look for non gravitational quantum field theories  satisfying these conditions. The necessary requirement is that as a consequence of energy self sourcing the total cross section  $\sigma_{tot}$ increases  with energy  $E$ as some positive power. 
  In this case the direct application of the optical theorem will lead to $s$ channel resonances,  {\it classicalons},  with life time of the order of $\sqrt{\sigma_{tot}}$ dominating the forward scattering amplitude. Again,  unitarity will imply  that these classicalons exhibit a degeneracy of the order $e^{aN}$ for some model-dependent coefficient $a$.

Since the energy self-sourcing translates into a coupling constants with dimensions of length, these theories are generically not UV complete in any Wilsonian sense. However the growth of $N$ with energy strongly indicates that the theory can classicalize and become complete and unitary. So how does  the notion of {\it classicality} fits within the Wilsonian paradigm ?

In Wilsonian sense a quantum field theory is defined as a {\it relevant} deformation of a UV CFT fixed point. For theories satisfying this condition the physics in the UV is completely characterized by the degrees of freedom defining the CFT fixed point. The problematic theories -- as gravity or Fermi theory-- are those defined at some scale by adding an {\it irrelevant} operator with a finite coupling. In this case when we flow into the UV the irrelevant coupling grows unbounded and the theory goes out of control. Normally UV completing these sort of theories -- within the frame of quantum field theory -- requires either to add new weakly-coupled degrees of freedom in such a way that the irrelevant coupling we started with can be obtained after integrating out  these new 
degrees of freedom  ($W$ bosons for the Fermi theory),  or changing the space-time scaling and consequently the nature of the interaction as irrelevant ( Horava-Lifshitz types of gravity ).  In a formal plane where the axis represents the relevant and the irrelevant operators  once we set the theory at a given scale, the RG flow into high energies will define a trajectory that for well-defined Wilsonian theories will end at a CFT  UV fixed-point,  while for ill-defined theories the trajectory will grow in the irrelevant direction . 

In order to understand classicality or self-UV-completion we need to improve the previous Wilsonian picture by adding a {\it new direction}  parametrized by $N(E)$. Note that $N(E)$ is defined as $N(E) = Er(E)$ with $r(E)$ the typical scale at energy $E$. In quantum mechanics this scale is the corresponding Compton (or de Broile) length and consequently $N(E)$ is {\it not flowing} with $E$ and stays of order one. However, in those cases where the theory generates a scale $r(E)$ that grows with $E$ and becomes larger than the Compton length in the deep UV,  we get a growing value of $N(E)$. In the case of gravity $r(E)$ is just the gravitational radius and the scale at which it becomes larger than the Compton length is at the unitarity bound of the theory, namely at the Planck length. Thus, until reaching Planckian energies, $N(E)$ remains constant and of order one. At ultra-Planckian energies $N(E)$ becomes larger than one and grows unbounded. Once we have added the $N$ direction what we see is that the theory, defined at some scale with an irrelevant coupling, remains in the {\it $N(E)\sim 1$ plane} until reaching the unitarity bound. Beyond the unitarity bound the theory starts to flow in the $N$ direction. 
Now the key point is, that in theories with growing $N(E)$ the UV flow is pushing the theory into the large $N$ region where the system becomes effectively classical. The growth of $N(E)$ only takes place in theories possessing a growing interaction scale $r(E)$ as it is the case in gravity. It is this RG flow into the large $N$ regime that  we characterize as self-UV-completion by classicalization. Models like Fermi theory remain in the $N=1$ plane and therefore cannot be self completed by classicalization. 
    
   This intrinsic gravitational flavor of any quantum field theory where energy acts as a source of the interaction is potentially very deep and interesting. t'Hooft's $\frac{1}{N}$ expansion \cite{tHooft} created a natural bridge between gauge theories and strings.   In order to put this in the context of our findings, what we have suggested in this paper can be understood as a {\it large $N$ approach} to gravity where the role of $N$ is played by the occupation number of gravitons and where the coupling is {\it self-tuned} to be order $\frac{1}{N}$ as a natural consequence of energy self-sourcing. In this frame large $N$ is not appearing as a simplifying approximation but as the unavoidable UV fate of energy self sourced theories. It is this quantum UV fate what we have baptized as classicalization.



    \vspace{5mm}
\centerline{\bf Acknowledgments}

 It  would like to thank Slava Mukhanov for very useful discussions and  comments. 
 We also thanks Alexey Boyarsky, Dieter Lust and Erik  Verlinde for discussions on some  aspects of 
 non-Wilsonian completion.  
   It is a pleasure to thank  Anton Zeilinger,  J\"org Schmiedmayer,  Philip Walther and other members of the Vienna  Institute for  Quantum Optics and Quantum Information for discussions on the analogies of the  phenomena discussed in this work with  non-gravitational quantum-mechanical systems and on prospects of simulating  these analogous systems in the lab-conditions, as well as   for  hospitality  during the visit of one of us.    
The work of G.D. was supported in part by Humboldt Foundation under Alexander von Humboldt Professorship,  by European Commission  under 
the ERC advanced grant 226371,  by European Commission  under 
the ERC advanced grant 226371,   by TRR 33 \textquotedblleft The Dark
Universe\textquotedblright\   and  by the NSF grant PHY-0758032. 
The work of C.G. was supported in part by Grants: FPA 2009-07908, CPAN (CSD2007-00042) and HEPHACOS P-ESP00346.


\begin{thebibliography}{99}

\bibitem{gia-cesar}
G.~ Dvali  and , C.~ Gomez, Self-Completeness of Einstein Gravity,  arXiv:1005.3497 [hep-th]; 

G. Dvali, S. Folkerts, C. Germani, ÒPhysics of Trans-Planckian GravityÓ, [ arXiv:1006.0984 [hep-th]]
Phys.Rev.D84:024039,2011. 



\bibitem{gia-gomez-alex}

G.~ Dvali, C.~ Gomez, Alex Kehagias.Classicalization of Gravitons and Goldstones.
arXiv:1103.5963 [hep-th],  JHEP 1111 (2011) 070




\bibitem{class}

G.~ Dvali, G.~ F. Giudice, C.~ Gomez, A.~ Kehagias,  UV-Completion by Classicalization, arXiv:1010.1415 [hep-ph]. JHEP 2011 (2011) 108; 

For recent discussions, see
C.~ Grojean,  R. S. Gupta, Theory and LHC Phenomenology of Classicalon Decays, 
 arXiv:1110.5317 [hep-ph], and references therein. 



\bibitem{no-hair} 

  W.~Israel,  {\it Phys. Rev.} {\bf 164} (1967) 1776;  {\it Commun. Math. Phys.}
{\bf 8}, (1968) 245;  

B.~ Carter, {\it Phys. Rev. Lett.}  {\bf 26}  (1971)  331.

J.~Hartle, {\it Phys. Rev.}  {\bf D 3} (1971) 2938.

  J.~Bekenstein, {\it Phys. Rev. }\  {\bf D 5}, 1239 (1972);
  {\it Phys.\ Rev.}\  {\bf  D 5},  (1972) 2403; {\it Phys. Rev. Lett.} {\bf 28} (1972) 452. 
  
  C.~Teitelboim, {\it Phys. Rev.} {\bf D 5} (1972) 294.  
 



\bibitem{species} 

G.~Dvali, ``Black Holes and Large N Species Solution to the
Hierarchy Problem,'' arXiv:0706.2050 [hep-th];

G.~Dvali and M.~Redi, ``Black Hole Bound on the Number of Species and Quantum Gravity at LHC,''
Phys. Rev.  {\bf D77} ( 2008) 045027,  arXiv:0710.4344 [hep-th];
G.~Dvali and C~Gomez, ``Quantum Information and Gravity Cutoff in Theories with Species'',
Phys. Lett. {\bf B674}  (2009) 303,  arXiv:0812.1940 [hep-th] . 










\bibitem{magneticbh}

 For an excellent review on magnetically-charged black holes, see, 
E.J.~Weinberg, http://arXiv: gr-qc/9503032. 

\bibitem{Witten}
E.~ Witten, 	
Baryons in the 1/n Expansion. Nucl.Phys. B160 (1979) 57

\bibitem{transplanck} G.~ Õt Hooft, Phys. Lett. B198, 61-63 (1987).

\bibitem{Veneziano}
D. Amati, M. Ciafaloni, G. Veneziano, 
Phys. Lett. B197, 81 (1987); Phys. Lett. B216, 41 (1989);Nucl. Phys. B347, 550-580 (1990).  Int. J. Mod. Phys. A3, 1615-1661 (1988).

\bibitem{Giddings}
S. B. Giddings, M. Schmidt-Sommerfeld, J. R. Andersen, ÒHigh energy scattering in 
gravity and supergravity,Ó Phys. Rev. D82, 104022 (2010). [arXiv:1005.5408 [hep-th]].


\bibitem{Bose}

See, e.g., F. Dalfovo et al., Rev. Mod. Phys. 71, 463 (1999)

\bibitem{tHooftSusskind} G .~'t Hooft, Dimensional Reduction in Quantum Gravity, Utrecht Preprint THU-93/26, 
gr-qc/9310006,
	
L. Susskind, The World as a hologram. J.Math.Phys. 36 (1995) 6377-6396, hep-th/9409089


\bibitem{tHooft} Gerard 't Hooft ,A Planar Diagram Theory for Strong Interactions.
 Nucl.Phys. B72 (1974) 461

\bibitem{Ads} J. M. Maldacena, Ò The Large N limit of superconformal Þeld theories and super- 
gravityÓ, Adv.Theor.Math.Phys. 2 (1998) 231, Int.J.Theor.Phys. 38 (1999) 1113, 
hep-th/9711200; 
S.S. Gubser, I.R. Klebanov, A. M. Polyakov, ÒGauge theory correlators from non- 
critical string theoryÓ, Phys. Lett. B428 (1998) 105, hep-th/9802109; 
E. Witten, ÒAnti-de Sitter space and holographyÓ, Adv.Theor.Math.Phys. 2 (1998) 
253, hep-th/9802150 









\end{thebibliography}
\end{document}